\documentclass[twoside,slac_one]{revtex4}
\usepackage{graphicx}
\usepackage{fancyhdr}
\usepackage{amsmath} 
\usepackage{bm}
\usepackage{amsxtra}
\usepackage{amssymb}
\usepackage{amsthm}
\usepackage{latexsym}
\usepackage{lscape}

\begin{document}

\title{Inclusive Search for Same-Sign Dileptons at ATLAS}

%

\author{B. Cerio on behalf of the ATLAS Collaboration}
\affiliation{Department of Physics, Duke University, Durham, NC, USA}
%

\begin{abstract}
An inclusive search for the non-Standard Model production of two isolated leptons of the same electrical charge is presented. The search was performed with 2010 collision data in the ATLAS detector at the Large Hadron Collider for $ee$, $e\mu$, and $\mu\mu$ channels. With 34 pb$^{-1}$ of integrated luminosity, no disagreement with the Standard Model prediction was found, and limits on the fiducial cross section for generic same-sign production as well as four model-dependent limits for models of new physics were set. An updated same sign analysis with 1.6 fb$^{-1}$ is underway and nearly complete for the $\mu\mu$ channel. With new event selection criteria, improved data-driven background estimates, and significantly more collision data, the sensitivity is expected to improve significantly beyond that of the analysis completed with 2010 data. Initial results from the updated analysis are presented. 

\end{abstract}

\maketitle

\thispagestyle{fancy}


\section{Introduction}
The production of two isolated, high-momentum leptons with the same electrical charge is rare in the Standard Model (SM). This signature, however, appears in a range of models beyond the SM, including supersymmetric models~\cite{Barnett:1993ea}, models with universal extra dimensions~\cite{Alwall:2008ag}, heavy Majorana neutrino models~\cite{Mohapatra:1974gc}, fourth-generation quark models~\cite{Frampton:1999xi}, as well as models with doubly-charged Higgs bosons~\cite{Perez:2008ha}. An inclusive search for same-sign (SS) dileptons, one in which only lepton kinematics are used to select the event, is therefore a promising testing ground for physics beyond the SM. 

Previous searches for SS dileptons have been done at Fermilab and the LHC. At Fermilab, the CDF collaboration searched inclusively, and found no evidence of physics beyond the SM~\cite{Abulencia:2007rd},~\cite{Aaltonen:2011rt}, while D0 performed a search for the associated production of the Higgs boson using SS dileptons~\cite{Abazov:2011ed}. Searches at CMS and ATLAS have generally been more signature-specific, placing requirements on missing transverse energy, jets, charged particles, total transverse energy, or an additional charged lepton~\cite{Aad:2011xm},~\cite{Chatrchyan:2011em}. In all of these analyses, no disagreement with the SM prediction was observed, and limits were placed on several models. 

In this note, we begin with a discussion of SS dilepton background sources. In section 3, the 2010 SS analysis is then briefly outlined, with a focus on the important results. We highlight the improvements made in the 2011 analysis in section 4.1, followed by a discussion of the analysis approach for the $\mu\mu$ channel, and presentation of the results. 

\section{SM Backgrounds}
In a SS dilepton analysis, understanding and estimating the backgrounds is the most challenging part, due to the fact that the dominant backgrounds are from mis-reconstructed objects. There are three sources of backgrounds from SM processes or mis-reconstruction: (1) diboson processes, including $ZZ$ and $WZ$, (2) SM processes in which two opposite sign leptons are produced and one of the leptons' charges is mis-measured, and (3) leptons from hadronic decays or charged hadrons themselves are reconstructed as prompt leptons. In addition to the SM contributions cited in (1), there are less significant contributions from $WWjj$ and $t\bar{t}W$. These ``real" SS backgrounds account for $\sim$17$\%$ of all backgrounds. Background contribution (2), called ``charge-flip", is primarily from Drell-Yan and top pair production. There are two ways in which the charge can be mis-measured. First, with high-$p_T$ leptons, the track is nearly straight, making it difficult to measure the curvature and thus the charge of the lepton. Second, in a pair of oppositely charged, high-$p_T$ electrons, if one of the electrons radiates a photon that subsequently converts to an asymmetric electron-positron pair, the three electron system from the prompt electron can be reconstructed as a prompt lepton of the same charge as the other prompt electron. This phenomenon, called ``trident electrons", is not found in the muon channel. The final background, (3), is by far the dominant one, accounting for $\sim$80$\%$ of the background to the SS signal. Quantum chromodynamic (QCD) dijet events produce these ``fake" same-sign leptons through heavy flavor quark decays. Additionally, $W$ + jet events that contain a lepton from heavy flavor decay as well as a prompt charged lepton from the $W$ boson decay can fake a true SS event at a non-negligible rate. 

\section{2010 Analysis}
With 34 pb$^{-1}$ of data integrated in 2010, an inclusive SS dilepton search was performed with three channels: $ee$, $e\mu$, and $\mu\mu$~\cite{Aad:2011vj}. Guided by the expected background components, the event selection was as follows. Both same sign lepton candidates were required to have a transverse momentum of greater than 20 GeV, and one of these lepton candidates was required to have triggered the event. Because the ATLAS tracking system covers up to a pseudorapidity of 2.5, leptons were required to fall between pseudorapidity values of -2.47 and 2.47. High quality tracks were obtained with hit requirements on the lepton track as well as a match between the charge as measured by the tracking system and the muon system, in the case of muons. A substantial portion of background from heavy flavor decays and charged hadron tracks was removed by rejecting leptons in which the transverse energy in the calorimeter within a cone of $\Delta{R}=0.2$ is greater than 15$\%$ of the lepton transverse energy. With lepton candidates selected, the dilepton system in the event is required to be SS and the lepton tracks are required to originate from the same primary vertex. The latter requirement suppresses contamination from cosmics, as well as pile-up phenomenon in which there is more than one collision in a bunch crossing. 

Background was estimated with both data-driven techniques and Monte Carlo simulation. For the fake background estimate, a data-driven technique was used to derive the rate at which non-prompt muons fake prompt muons. This fake rate was then applied to the appropriate data distributions, yielding an estimate of the fake component of the background. For the charge-flip estimate, a combination of data and MC was used in the $Z$ peak control region, yielding a prediction of zero events for this dataset. Finally, MC was used to predict the diboson background component with the uncertainty in the cross-section being accounted for with a systematic. 

\begin{figure}[h]
\centering
\includegraphics[scale=0.8]{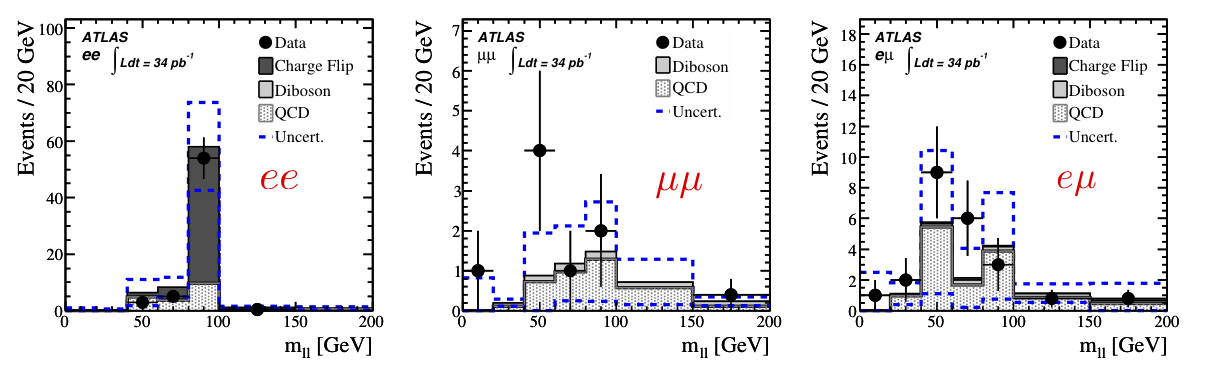}
\caption{Invariant mass distributions of all three channels in the 2010 analysis.} \label{invarMass2010}
\end{figure}

\begin{figure}[h]
\centering
\includegraphics[scale=0.8]{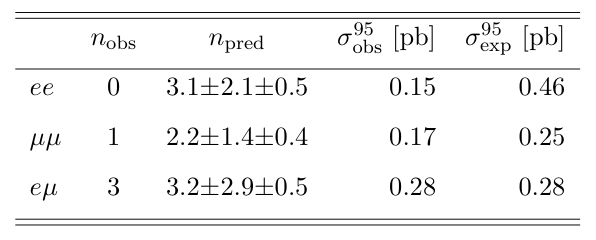}
\caption{The observed and predicted yields of same-sign dilepton events in the phase space region $M_{ll}>110$ GeV. Fiducial cross section limits are also displayed.} \label{fidCrossSecLim}
\end{figure}

\begin{figure}[h]
\centering
\includegraphics[scale=0.5]{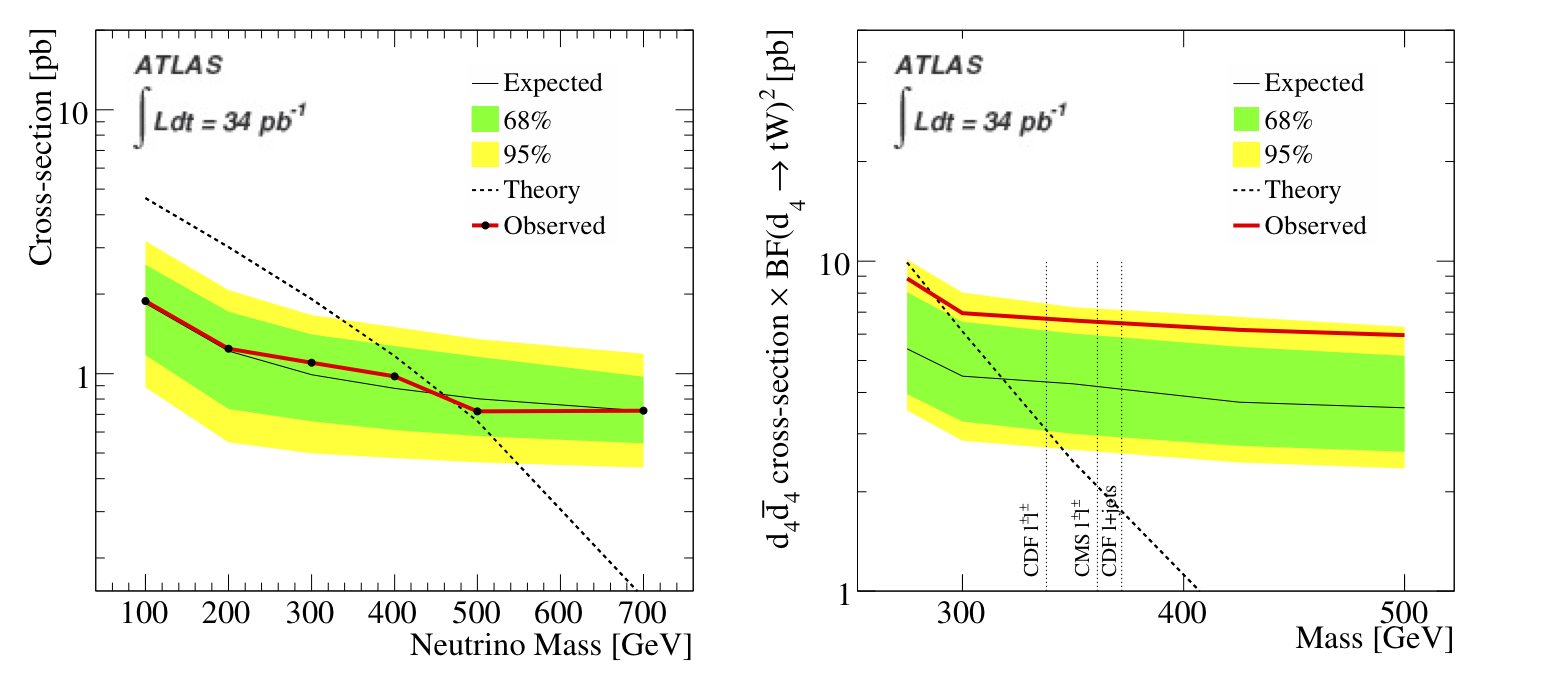}
\caption{Predicted and observed upper limits on cross sections for production of (a) heavy Majorana neutrinos and (b) fourth generation $d$ quarks. The theoretical cross sections are overlaid, and the exclusion regions can be seen.} \label{crossSecUpperLimit2Model}
\end{figure}

The resulting invariant mass spectrum of the dilepton system after the above cuts is shown in ~Figure~\ref{invarMass2010}. There is agreement between the SM prediction and the observed event yield in all channels, allowing fiducial cross section limits on the production of same-sign lepton pairs to be set. These limits, as well as the predicted and observed event yields, are displayed in ~Figure~\ref{fidCrossSecLim} for the region $M_{ll}>110$ GeV. Additionally, cross section upper limits were set in four specific models, a heavy Majorana neutrino model, models with lepton cascades, a L-R symmetric model, and a fourth generation $d$ quark model. In ~Figure~\ref{crossSecUpperLimit2Model}, the observed and predicted cross section upper limits are shown as a function of model parameters with model-specific cross sections overlaid. Referring to ~Figure~\ref{crossSecUpperLimit2Model}(a), heavy Majorana neutrinos below a mass of $\sim$450 GeV are excluded, assuming that these heavy neutrinos are produced with by a four fermion operator and the scale of new physics, $\Lambda$, is on the order of one TeV. In ~Figure~\ref{crossSecUpperLimit2Model}(b) the upper limit is plotted as a function of the mass of the fourth generation $d$-type quark, and the exclusion region is set below a mass of 320 GeV using the pair production channel $d_4\bar{d}_4\rightarrow{tW\bar{t}W}$ with the assumption that the branching ratio of this heavy $d$ quark decay is 100$\%$. 

\section{2011 analysis}

The 2011 SS dilepton analysis is currently underway. Because the $\mu\mu$ channel is near completion, the remainder of the note will focus on the details of this channel~\cite{Aad:2011bc}. 

\subsection{Analysis improvements}

Beyond the obvious improvement of a factor of $\sim$50 more collision data, there are several improvements to the 2011 SS dimuon analysis. First, the $p_T$ cut on the subleading muon was lowered to 10 GeV, yielding greater acceptance for potential signals. Second, the isolation requirement was changed from a requirement on the calorimeter energy deposited in a cone around a lepton candidate to a requirement on the sum of the track $p_T$ in a cone around the candidate. This track-based isolation cut has been shown to have less dependence on pileup, which is becoming more of a consideration with an average of six interactions per beam crossing in the first fb$^{-1}$ integrated in 2011. Finally, there is a cut on the impact parameter significance, a cut that has been shown to be $\sim$99$\%$ efficient for prompt muons, while suppressing background from $b$ hadron decays by a factor of three. 

\subsection{Background estimation}

\begin{figure}[h]
\centering
\includegraphics[scale=1]{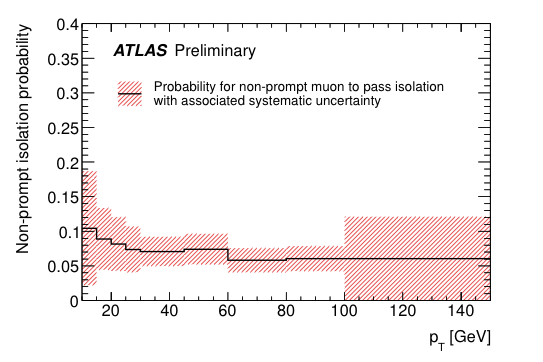}
\caption{Rate at which muons from hadronic sources fake prompt muons as a function of muon $p_T$.} \label{fakeRate}
\end{figure}

\begin{figure}[h]
\centering
\includegraphics[scale=0.75]{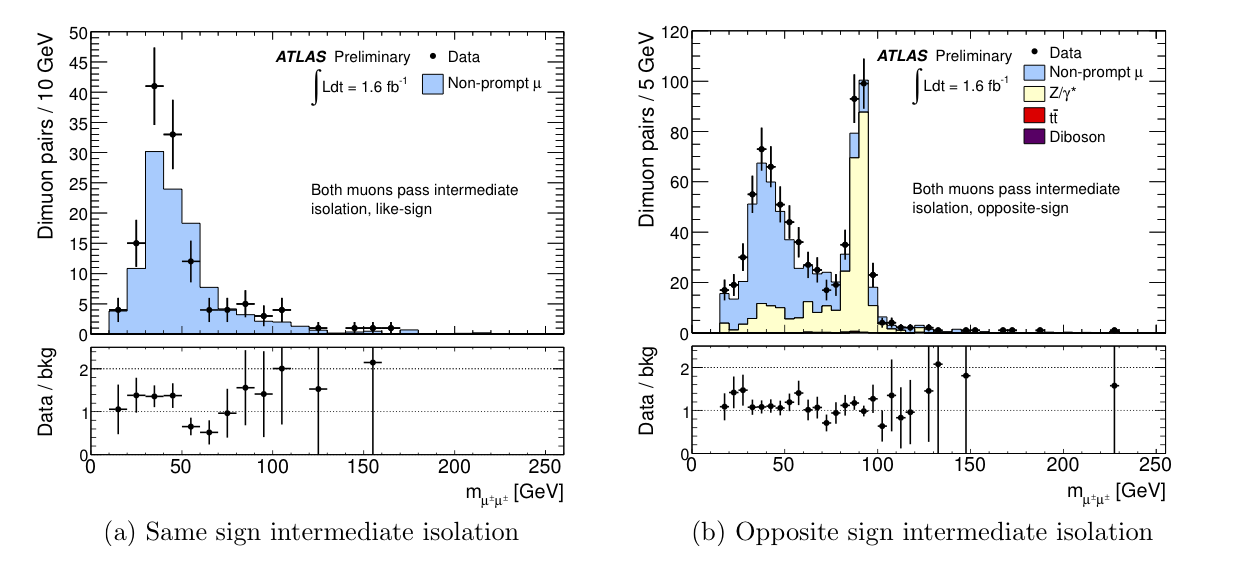}
\caption{Dimuon invariant mass distribution in intermediate isolation control region for (a) same-sign muons and (b) opposite sign muons.} \label{controlRegionM}
\end{figure}

In the muon channel non-prompt muons from semi-leptonic $b$ hadron decays and pion or kaon decays-in-flight account for 83$\%$ of the SS background. In order to estimate this background, a data driven technique similar to that of the 2010 analysis was applied to obtain the event yield for fakes. The fake rate in this analysis was defined as the fraction of non-prompt muons that pass the isolation requirement among those passing the other selection requirements. Three heavy-flavor enriched control samples were selected to obtain the fake rate displayed in ~Figure~\ref{fakeRate}. Differences in the fake rate obtained from each control sample were used to quantify the systematic uncertainty in each $p_T$ bin. There was at least a 30$\%$ uncertainty for each bin with the highest uncertainty of 80$\%$ in the lowest $p_T$ bins. The fake rate above 100 GeV was approximately flat at 7$\%$. In the implementation of the fake rate to obtain an event yield, the rate is binned in $\eta$ in addition to $p_T$. To validate the fake rate method, a fake-enhanced control region was considered. In one such control region, an intermediate isolation requirement was placed on the muons, and then the fake rate method was applied, yielding the result in ~Figure~\ref{controlRegionM}-- the invariant mass distributions of (a) same-sign and (b) opposite-sign muon pairs. For the same-sign control region 112 events were predicted and 129 were observed, while for the opposite sign region, 339 were predicted and 391 observed. These observations agree with the fake-rate-method-derived prediction within the systematic uncertainty associated with the fake rate. 

The charge-flip background estimate was obtained from a data-driven upper limit on the charge flip rate, and was found to be negligible for this particular dataset. As in the 2010 analysis, the diboson background contribution was obtained from MC. 

\subsection{Results}

\begin{figure}[h]
\centering
\includegraphics[scale=1]{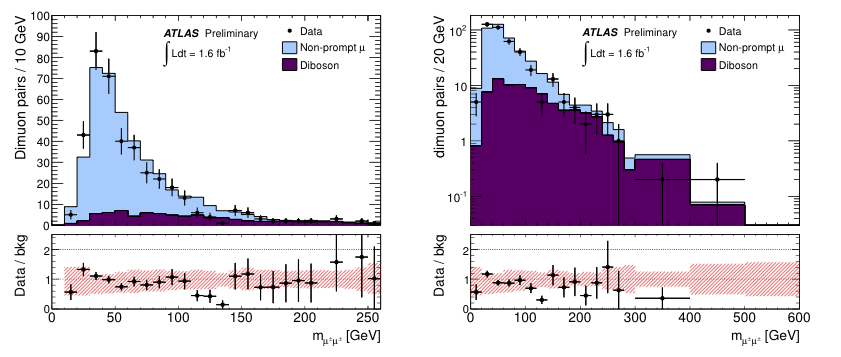}
\caption{Background prediction for the SS dimuon invariant mass spectra on log and linear scales.} \label{invarMass2011}
\end{figure}

\begin{figure}[h]
\centering
\includegraphics[scale=0.5]{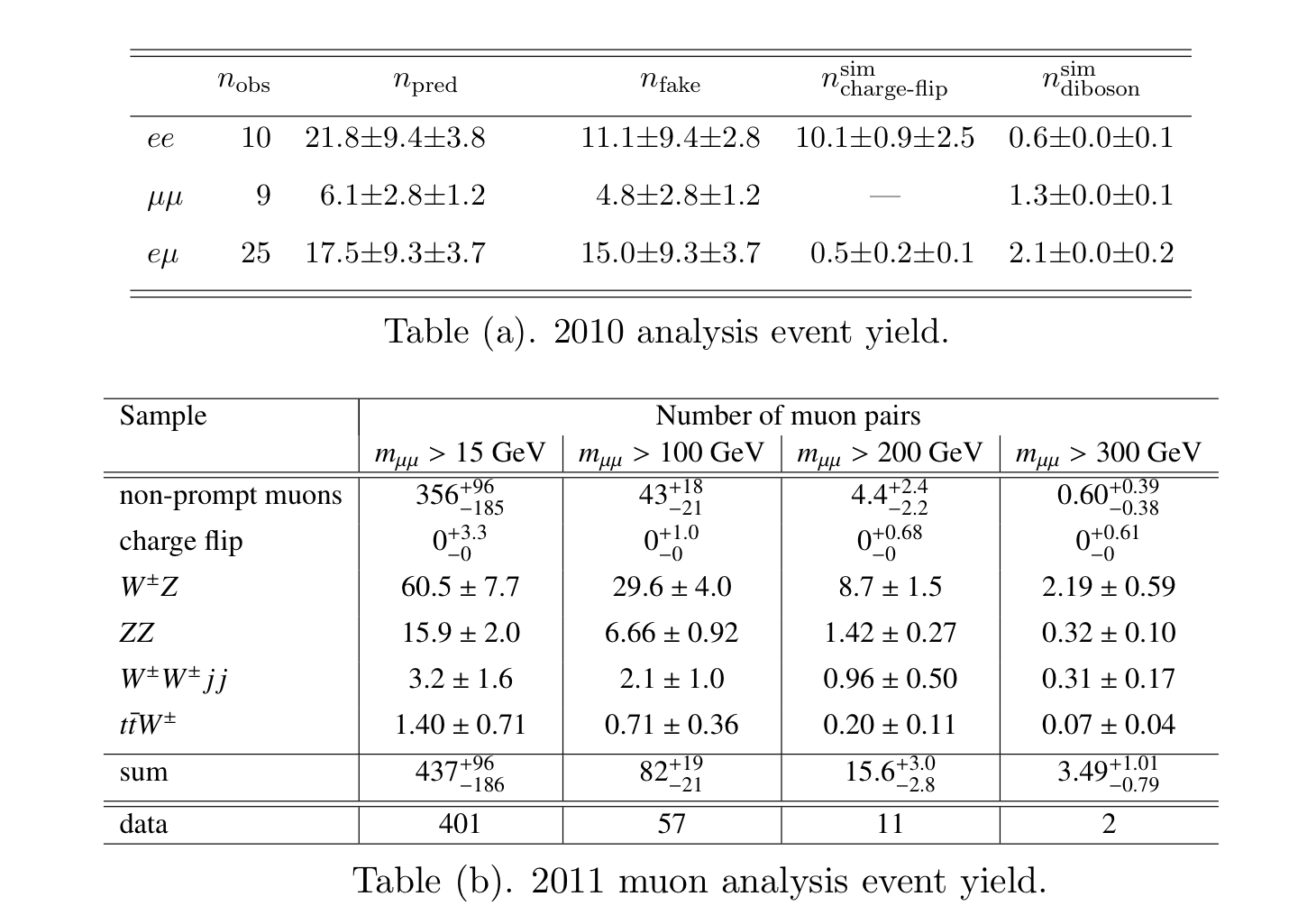}
\caption{Predicted and observed event yields for (a) the 2010 analysis and (b) the 2011 dimuon analysis.} \label{eventYield2011}
\end{figure}

With a validated method for estimating the background, the following results were obtained for the dimuon channel. In ~Figure~\ref{invarMass2011}, the invariant mass for SS dileptons is plotted on linear and log scales. The non-prompt background is dominant with a small contribution from dibosons. Since the charge-flip rate for muons is small, there is no charge-flip contribution in this dataset. The expected event yields are shown in ~Figure~\ref{eventYield2011}. In 1.6 fb$^{-1}$, 401 events were observed, a value that agrees with the SM prediction of 437$^{+96}_{-186}$ in the region $M_{\mu\mu}>15$ GeV. In the 2010 analysis, the prediction was 6.1 events and 9 were observed outside of the $Z$ mass window. We therefore expect significantly greater sensitivity in the 2011 SS dimuon analysis. 

\section{Conclusion}

An inclusive search for SS dileptons was performed with 32 pb$^{-1}$ of data recorded in 2010 in the ATLAS detector at the LHC, and a similar search is in progress with at least 1.6 fb$^{-1}$ of data integrated in 2011. With the 2010 dataset, no excess beyond the SM prediction was observed, and competitive limits were placed on four models of new physics, including a heavy Majorana neutrino model and a fourth generation $d$-type quark model. With a factor of almost 50 more data, and several selection and background estimation improvements, it is expected that limits obtained with the 2011 dataset will be $\emph{much}$ more stringent if no disagreement with the SM is found.


\bigskip 

\end{document}